\documentclass[12pt]{iopart}
\usepackage{graphicx}  
\begin{document}

\title[Weak universality of spin-glass transition]
{
Weak universality of spin-glass transitions\\ in three-dimensional
$\pm J$ models
}

\author{Tota Nakamura, Shin-ichi Endoh
\footnote[1]{Present address:
BIRDS SYSTEMS RESEARCH INSTITUTE INC.,
1-6-15 Hirakawa-cho, Chiyoda-ku, Tokyo, 102-0093, Japan
}
 and Takeo Yamamoto
}

\address{
Department of Applied Physics, Tohoku University, \\ Aoba-yama 05,
Sendai, Miyagi, 980-8579, Japan
}

%\Large

%\baselineskip 34pt

\begin{abstract}
We find a possibility of a weak universality of spin-glass
phase transitions in three-dimensional $\pm J$ models.
The Ising, the XY and the Heisenberg models
seem to undergo finite-temperature phase transitions with a ratio of
the critical exponents $\gamma/\nu \sim 2.4$.
Evaluated critical exponents may explain corresponding experimental results.
The analyses are based upon nonequilibrium relaxation from a
paramagnetic state and finite-time scaling.
\
\end{abstract}



%\maketitle

\section{Introduction}
A spin-glass (SG) phenomenon has been attracting great interest
both theoretically and experimentally \cite{sgreview}.
Applications now cover a wide range of interdisciplinary fields of statistical
physics and informational physics, as treated in this special issue.
However, many subjects are not well understood, in spite of 
efforts made over almost thirty years.
One of these subjects is whether or not the SG transition of real materials
can be explained by a simple random-bond spin model.

Spins of many SG materials are well-approximated by the Heisenberg spins.
The simplest theoretical model is the Heisenberg model with
random nearest-neighbour interactions.
However, numerical studies suggest that there is no
finite-temperature SG transition in this model \cite{mcmillan,olive}.
Kawamura \cite{chirality1,chirality2} proposed the chirality mechanism
in order to solve this discrepancy.
The chiral-glass (CG) transition occurs without the SG order.
A small but finite random anisotropy in the real materials
mixes the chirality degrees of freedom and the spin degrees of freedom.
This anisotropy effect induces the SG transition observed in the real materials.
The scenario is based upon results that the SG transition does not 
occur in the isotropic model.
However, Matsubara et al. \cite{matsubara1,matsubara2,matsubara3} 
recalculated the domain-wall excess energy and the SG susceptibility, from 
which they suggested that the finite-temperature SG transition 
does possibly occur.
Methods are quite similar to the previous ones \cite{mcmillan,olive}.
Subtle differences in the analyses of the obtained data drew an 
opposite conclusion.

The spin-glass problem is one of the most difficult subjects in
computational physics.
It can be a tough bench-mark test for a new numerical method.
It may be applied to other complex systems, if successful in the spin-glass
investigations.
The difficulty is caused by serious slow dynamics.
It requires many Monte Carlo steps to reach the equilibrium states.
An observed quantity at each step has a strong correlation
even after the equilibration.
The system sizes which can be treated in the simulations are accordingly 
limited to very small ones, 
e.g.,  mostly a linear size is twenty or less in three dimensions.
Size effects are generally stronger in the continuous spin systems because
the spins are soft and the boundary effect propagates faster.
Frustration and randomness also yield a considerable size effect.
The system sizes treated previously in the studies of the Heisenberg
SG models are too small to extrapolate to the thermodynamic limit.
This is our motivation for reexamining the SG transition using the 
nonequilibrium relaxation (NER) method\cite{ner1,ner2,ner3,ner4,huse,blundell}.

\begin{figure}[ht]
\begin{center}
\includegraphics[width=7cm]{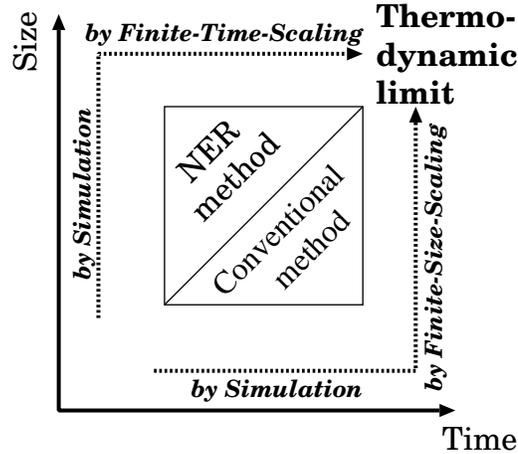}
\caption[]{A schematic diagram to approach the thermodynamic limit.}
\label{fig:gainen}
\end  {center}
\end{figure}

The difficulty mentioned above can be overcome by using the NER method.
This method takes an opposite approach to the thermodynamic limit.
\Fref{fig:gainen} schematically shows a comparison between the
conventional equilibrium simulational method and the NER method.
In the conventional method one takes the infinite time limit first by
achieving the equilibrium states in finite sizes.
The thermodynamic limit is taken by the finite-size scaling analysis of the
obtained data.
In the NER method we take the infinite size limit first by dealing with a very
large system within a finite time range before the finite-size effect appears.
Then, the {\it finite-time scaling analysis} \cite{timescaling1,timescaling2}
is performed to obtain the thermodynamic properties.
The cost of a simulation is in the same order of $L^{d+z}$ for both methods.
However, a coefficient factor in the NER method is much smaller 
than that in the conventional method.
An observation time length in the NER method is sufficient if we can 
observe a beginning of a final relaxation to the equilibrium states 
(equilibrium relaxation).
On the other hand, it is necessary to wait until the end of 
the equilibrium relaxation in the conventional method.
The latter time scale is typically $10 \sim 10^2$ times longer than
the former one in the spin glass models.
(For example, $\chi_\mathrm{sg}$ of $L=17$ in \fref{fig:pofq} (a) or
$\chi_\mathrm{sg}$ at $T=0.56$ in \fref{fig:xy} (a).)
Therefore, the NER method has an advantage over the conventional method
by this factor.
We use the residual computational time to enlarge the system size
and to increase statistical accuracy.

By using the NER method we have made it clear 
that the SG transition occurs in the
Heisenberg model at the same finite temperature as the CG transition occurs 
\cite{totasg1}.
The estimated critical exponent $\gamma$ is consistent with the
corresponding experimental result \cite{heisenexp}.
The chirality mechanism is not necessary to explain the spin-glass experiments
since the chirality trivially freezes if the spin freezes.
However, one may question the use and the validity 
of the NER method in the spin-glass phenomenon.
Therefore, we have corroborated our method by studying the Ising SG model.
Many numerical investigations 
\cite{ogielski,bhatt,kawashimayoung,palassini,maricampbell}
yield consistent results on the existence of the SG transition, 
the critical temperature and the critical exponents.
They are also consistent with the
corresponding experimental results \cite{isingsgexp}.
The NER method yields consistent results for a small number of
simulations as discussed in \sref{sec:results}.

In this procedure we have found a possibility of a weak universality: 
a critical exponent divided by $\nu$, for example $\gamma/\nu$,
is common among models in a weak universality class.
A ratio of the critical exponents $\gamma/z\nu$ appearing in a finite-time 
scaling analysis is found to be consistent between the Heisenberg model
and the Ising model.
We have verified that a ratio $\gamma/\nu$ is also consistent by
evaluating the dynamic exponent $z$ alone.
The analysis is expanded to the XY SG model and the value is also 
found to be consistent.
These findings are quite surprising.
We must reconsider the role of the spin dimensions 
and the distribution of the randomness in the SG phase transition.

This paper is organised as follows:
In \sref{sec:model} the model and the method are explained.
Descriptions of the procedure of the NER method 
and the finite-time scaling are given.
In \sref{sec:results} the results on the Ising model, the Heisenberg model and
the XY model are shown. 
Then, the possibility of a weak universality is discussed.
\Sref{sec:summary} is devoted to a summary.

%%%%%%%%%%%%%%%%%%%%%%%%%%%
\section{Model and Method - Nonequilibrium relaxation}
\label{sec:model}
%%%%%%%%%%%%%%%%%%%%%%%%%%%

A model treated in this paper
is the nearest-neighbour $\pm J$ random-bond model,
\begin{equation}
{\cal H}=\sum_{\langle i,j \rangle} J_{ij} 
\mbox{\boldmath $S$}_i \cdot \mbox{\boldmath $S$}_j.
\end  {equation}
A linear size of a lattice is denoted by $L$.
Skewed periodic boundary conditions are imposed, i.e., 
total numbers of spins $N=L\times L\times (L+1)$.
An interaction
$J_{ij}$ takes two values of $+J$ and $-J$ with the same probability.
The temperature $T$ is scaled by $J$.

Spins are updated by a single-spin-flip algorithm.
The Metropolis (M) update is used in all models, whereas
the heat-bath (H) update \cite{olive} is used in the Heisenberg model.
Physical quantities observed in our simulations are the SG
susceptibility $\chi_\mathrm{sg}$, the CG susceptibility
$\chi_\mathrm{cg}$ and the Binder parameter in regard to 
the spin-glass transition $g_\mathrm{sg}$.
These quantities are calculated through the overlap between real replicas.

First, we rewrite the thermal average by an arithmetic mean over 
thermally equilibrium ensembles labelled by $\alpha$ as
\begin{equation}
\langle \mbox{\boldmath $S$}_i \cdot \mbox{\boldmath $S$}_j \rangle
= \frac{1}{m}\sum_{\alpha=1}^{m} 
\mbox{\boldmath $S$}_i^{(\alpha)}
\cdot
\mbox{\boldmath $S$}_j^{(\alpha)}
= \frac{1}{m}\sum_{\alpha=1}^{m} 
\sum_{\mu}^{x,y,z}
S_{i,\mu}^{(\alpha)} S_{j,\mu}^{(\alpha)}.
\label{eq:thermalreplica}
\end  {equation}
The bracket $\langle \cdots \rangle$ denotes the thermal average and 
$m$ denotes a number of ensembles.
The index $\mu$ stands for three components of spins: $x, y$ and $z$.
This expression is substituted into the definition of the SG susceptibility:
\begin{equation}
\chi_\mathrm{sg} = 
\frac{1}{N}\sum_{i,j}\left[
\langle \mbox{\boldmath $S$}_i \cdot \mbox{\boldmath $S$}_j 
\rangle^2
\right]_\mathrm{c}
= N \left[
\frac{1}{m^2}\sum_{\alpha, \beta}^{m}\sum_{\mu,\nu} ^{x,y,z}
(q^{\alpha\beta}_{\mu,\nu})^2
\right 
] _\mathrm{c},
\end  {equation}
where
$q_{\mu, \nu}^{\alpha \beta}
\equiv (1/N)\sum_i S_{i, \mu}^{(\alpha)} S_{i, \nu}^{(\beta)}$
is an overlap between the $\mu$ component of a spin $i$ 
on an ensemble $\alpha$: $S_{i, \mu}^{(\alpha)}$ and the $\nu$ component of
the spin on an ensemble $\beta$: $S_{i, \nu}^{(\beta)}$.
The bracket $[\cdots ]_{\rm c}$ denotes the configurational average.

Here, we introduce the following real replicas.
Each real replica takes the same random bond configuration and the
different paramagnetic initial spin state.
They are updated in parallel with different random number sequences.
This procedure corresponds to quenching from an infinite temperature.
The thermal ensembles are realized by these real replicas which 
approach different equilibrium states.
Therefore, we replace the thermal average by the average over these
real replicas as equation \eref{eq:thermalreplica}.
The indices $\alpha$ and $\beta$ now represent real replicas.
We do not take into consideration a constant term which arises 
from the overlap between the same
replica $\alpha=\beta$ and use the following expressions in the simulations.
\begin{eqnarray}
\chi_\mathrm{sg}&=& N \left [
\frac{2}{m(m-1)}\sum_{\alpha > \beta}^{m}\sum_{\mu,\nu} ^{x,y,z}
(q^{\alpha\beta}_{\mu,\nu})^2 \right ] _\mathrm{c},
\\
\chi_{\rm cg}&=&
\frac{1}{3N} \left[\frac{2}{m(m-1)}\sum_{\alpha > \beta}^m \left(
\sum_{i, \phi}
C_{i, \phi}^{(\alpha)}
C_{i, \phi}^{(\beta)}\right)^2\right]_{\rm c}
,
\\
g_\mathrm{sg}&=&\frac{1}{2}\left(
A-B\frac{\displaystyle \sum_{\mu,\nu,\delta,\rho}\left[
\frac{2}{m(m-1)}\sum_{\alpha > \beta}^m
(q^{\alpha\beta}_{\mu,\nu})^2 (q^{\alpha\beta}_{\delta,\rho})^2
\right]_{\rm c}
}
{\displaystyle \left(\sum_{\mu,\nu}\left[
\frac{2}{m(m-1)}\sum_{\alpha > \beta}^m
(q^{\alpha\beta}_{\mu,\nu})^2
\right]_{\rm c}\right)^2}
\right).
\end  {eqnarray}
A number of replicas $m$ controls the precision of the thermal average.
It is better to take a large value.
We prepare eight or nine replicas for each bond configuration in this paper.
The scalar chirality is defined by three neighbouring spins as
$
C_{i, \phi}^{(\alpha)}=
\mbox{\boldmath $S$}_{i+\hat{\mbox{\boldmath $e$}}_{\phi}}^{(\alpha)}
\cdot
(
\mbox{\boldmath $S$}_{i}^{(\alpha)}
\times
\mbox{\boldmath $S$}_{i-\hat{\mbox{\boldmath $e$}}_{\phi}}^{(\alpha)}
),
$
where $\hat{\mbox{\boldmath $e$}}_{\phi}$ denotes a unit lattice vector along
the $\phi$ axis.
In the XY model we calculate the vector chirality, which is defined by
$
C_{i,\phi}^{(\alpha)}=(1/2\sqrt{2})(
J_{ij} 
\mbox{\boldmath $S$}_{i}^{(\alpha)} \times \mbox{\boldmath $S$}_{j}^{(\alpha)}
+J_{jk}
\mbox{\boldmath $S$}_{j}^{(\alpha)} \times \mbox{\boldmath $S$}_{k}^{(\alpha)}
+J_{kl}
\mbox{\boldmath $S$}_{k}^{(\alpha)} \times \mbox{\boldmath $S$}_{l}^{(\alpha)}
+J_{li}
\mbox{\boldmath $S$}_{l}^{(\alpha)} \times \mbox{\boldmath $S$}_{i}^{(\alpha)}
)|_z.
$
Indices $i,j,k,l$ denote four sites forming a square plaquette in the $\phi$
direction from the $i$ site.
Constants in a definition of $g_\mathrm{sg}$ are:
$A=3, B=1$ for the Ising model, $A=6, B=4$ for the XY model and
$A=11, B=9$ for the Heisenberg model.

\begin{figure}[ht]
\begin{center}
\includegraphics[width=15cm]{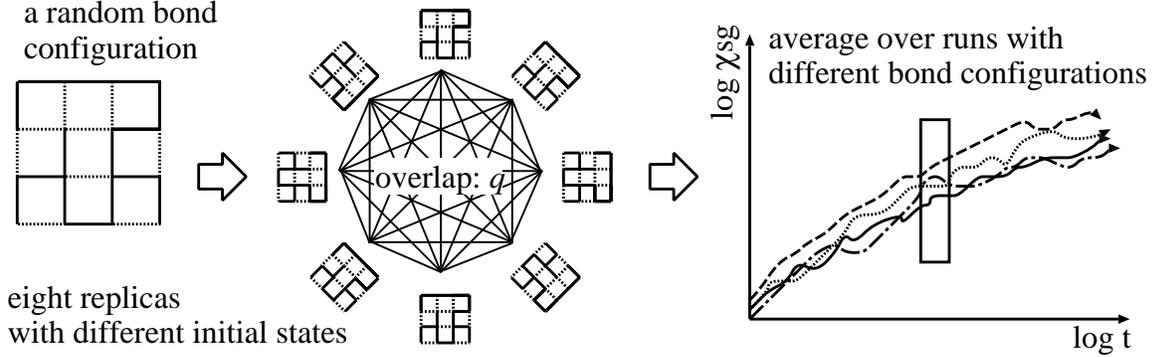}
\caption[]{A schematic flow of our simulation.
Solid bonds and broken bonds in the lattice
depict ferromagnetic bonds and antiferromagnetic bonds.}
\label{fig:method}
\end  {center}
\end{figure}

\Fref{fig:method} shows a schematic diagram of the simulation procedure.
We calculate a physical quantity at each time step $t$ and obtain a
relaxation function.
Another simulation starts by changing a random bond configuration, initial
spin states and a random number sequence.
Then, another relaxation function is obtained.
Finally, we take an average of data at each step 
over these different Monte Carlo runs.
It should be noted that the average is over independent data.
It guarantees an absence of systematic error due to correlations of the 
observed quantity, which we usually encounter in the conventional Monte Carlo 
time average.
The obtained raw relaxation function is utilised by the following
finite-time scaling analysis.

The most important point in the NER method is to exclude the finite-size 
effect from the raw relaxation function.
The method is based upon taking the infinite-size limit first.
If a relaxation function includes a finite-size effect,
it exhibits converging behaviour 
because every finite system has a definite equilibrium state.
This behaviour misleads us into thinking that 
the temperature is in the paramagnetic phase 
even though it is the critical temperature.
Therefore, the critical temperature is always underestimated if the 
size is insufficient.
We check the size effect by changing the lattice sizes and 
always confirm a time range in which the size can be considered as infinity.

The SG susceptibility is expected to diverge at the critical temperature 
($T_\mathrm{sg}$) as $\chi_\mathrm{sg}(t)\sim t^{\gamma/z\nu}$\cite{huse}.
We obtain $T_\mathrm{sg}$, $\gamma$ and $z\nu$ by the finite-time scaling 
analysis on the relaxation functions of $\chi_\mathrm{sg}(t)$
in the paramagnetic phase.\cite{totasg1}
Since the initial spin configuration is completely random,
$\chi_\mathrm{sg}(t=0)\sim 1$.
We start a set of simulations at a temperature $T$ that is obviously in the
paramagnetic phase.
The relaxation function $\chi_\mathrm{sg}(t)$ at this temperature increases
with $t$ but soon converges to a finite value.
As the temperature is lowered to approach the critical temperature,
the relaxation function tends to show diverging behaviour.
Since the temperature is still in the paramagnetic phase, the relaxation 
finally converges to a finite value after a correlation time $\tau(T)$.
The spin-glass correlation increases with time and reaches the correlation 
length $\xi(T)$ after this correlation time.
Two quantities relate with each other by $z$ as $\tau(T)\sim \xi^z(T)$.
Therefore, the correlation time should diverge at $T_\mathrm{sg}$ as
\begin{equation}
\tau(T)\sim (T-T_\mathrm{sg})^{-z\nu}.
\label{eq:taufit}
\end  {equation}
The correlation time can be estimated by scaling the raw relaxation function.
We obtain $\gamma/z\nu$ and $\tau(T)$ so that the scaled functions
$\chi_\mathrm{sg}(t) t^{-\gamma/z\nu}$
at all temperatures plotted against  $t/\tau(T)$ fall onto a single curve.
Then, the critical temperature and the exponent $z\nu$ 
are estimated by the least-squares fitting with the equation \eref{eq:taufit}.
Since a ratio $\gamma/z\nu$ is already estimated by the scaling,
$\gamma$ is obtained.

The NER of the Binder parameter $g_\mathrm{sg}(t)$ is calculated at the
obtained $T_\mathrm{sg}$.
Since quantity is related to the fourth-order cumulant,
many bond samples are necessary to obtain meaningful data.
The number of bond configurations to obtain the results in this paper
is summarised in \tref{tab:samplist}.
The Binder parameter is expected to diverge at $T_\mathrm{sg}$
as $g_\mathrm{sg}(t)\times L^d \sim t^{d/z}$\cite{blundell}, by which
$z$ is independently obtained.
Then, $\nu$ is estimated from a value of $z\nu$ obtained by the $\tau$-fitting
explained above.
All exponents are now estimated by the scaling relation.
It is possible to compare the critical exponents with the experimental results.

The last procedure of our method is to corroborate the results
by observing the NER of $\chi_\mathrm{sg}$ at the obtained $T_\mathrm{sg}$.
It should diverge as $t^{\gamma/z\nu}$ 
with the same exponent obtained by finite-time scaling.
If the exponents are inconsistent, the scaling analysis is misled by 
an insufficient time range or by the finite-size effect.
In the Ising model we perform another check at $T_\mathrm{sg}$ by observing 
the NER of the distribution function of the replica overlap, $P(q,t)$.
The finite-time scaling plot of $P(q,t)$ should ride on a single scaling
function with the same exponent obtained by finite-time scaling 
of $\chi_\mathrm{sg}$.
This is a direct interpretation of the finite-size scaling of
$P(q,L)$ \cite{bhatt} by $t\propto L^z$.

\section{Results}
\label{sec:results}

Numbers of bond configurations to obtain data at the critical temperature
are summarised in \tref{tab:samplist}.
The numbers at other temperatures are mostly in the same order.
For each bond configuration, we prepared eight replicas for the XY and the 
Heisenberg model and nine replicas for the Ising model.

\begin{table}[ht]
\centering
\caption{Numbers of bond configurations to obtain data of $\chi_\mathrm{sg}$
and $g_\mathrm{sg}$ at $T_\mathrm{sg}$ in this paper.
Indices (M) and (H) in the Heisenberg model denote update algorithms:
(M) for the Metropolis and (H) for the heat-bath.
Arrows mean that the number is same as to the right.}
\begin{tabular}{clc|rrrrrrr}
\hline
&&&Step&&&&&&\\
&Model&Size& $10^3$&$10^4$&$5 \cdot 10^4$&$10^5$
&$5\cdot 10^5$&$10^6$&$4\cdot 10^6$\\
\hline
$\chi_\mathrm{sg}$& Ising&49&$\to$&$\to$&$\to$&393&$\to$&$\to$&88\\
& XY&39&$\to$&$\to$&$\to$&5246&120&&\\
& Heisenberg(M)&59&$\to$&$\to$&$\to$&$\to$&$\to$&104&\\
& Heisenberg(H)&89&$\to$&58&$\to$&22&&&\\
\hline
$g_\mathrm{sg}$& Ising&39&255480&85480&18576&12626&$\to$&1830&172\\
& XY       &19&$\to$&$\to$&$\to$&7803&&&\\
& Heisenberg(H)&39&43114&18316&7038&&&&\\
\hline
\end{tabular}
\label{tab:samplist}
\end{table}

\subsection{Ising model}

\begin{figure}[ht]
\centering
\includegraphics[width=7.5cm]{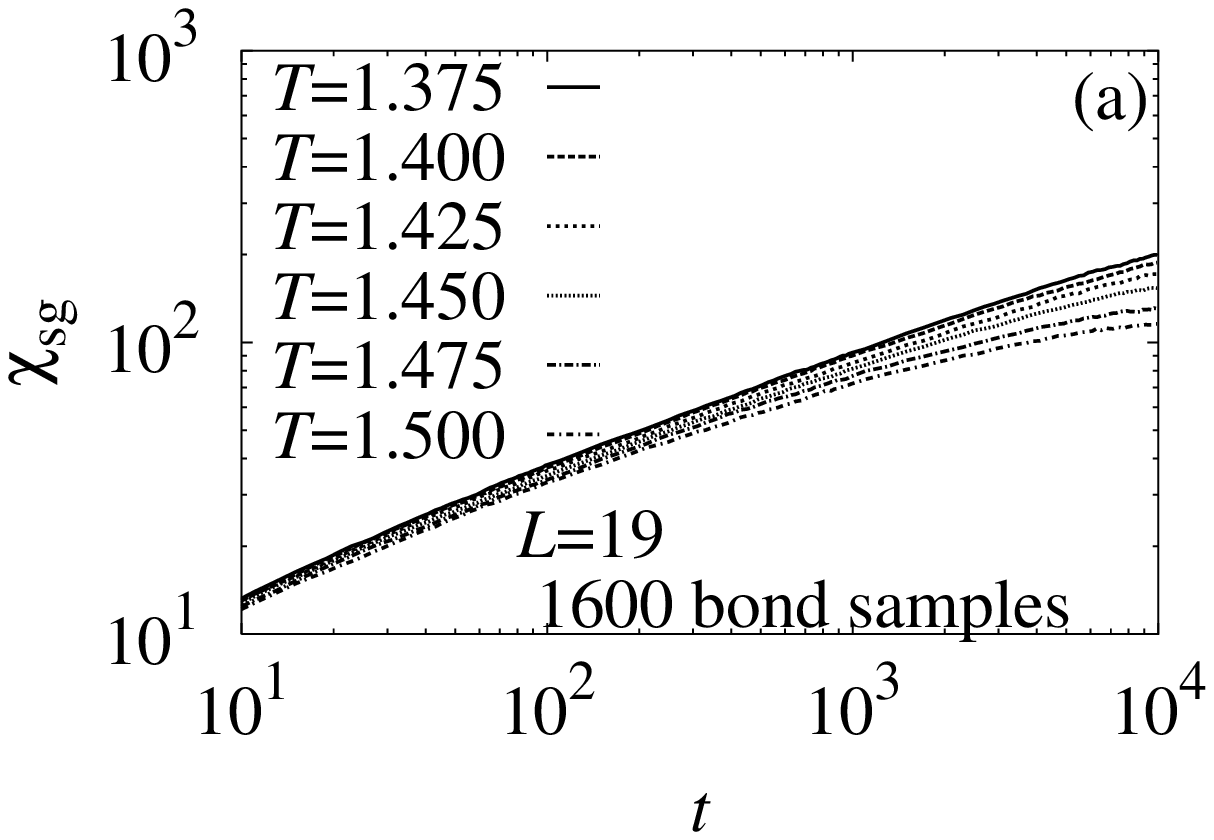}
\includegraphics[width=7.5cm]{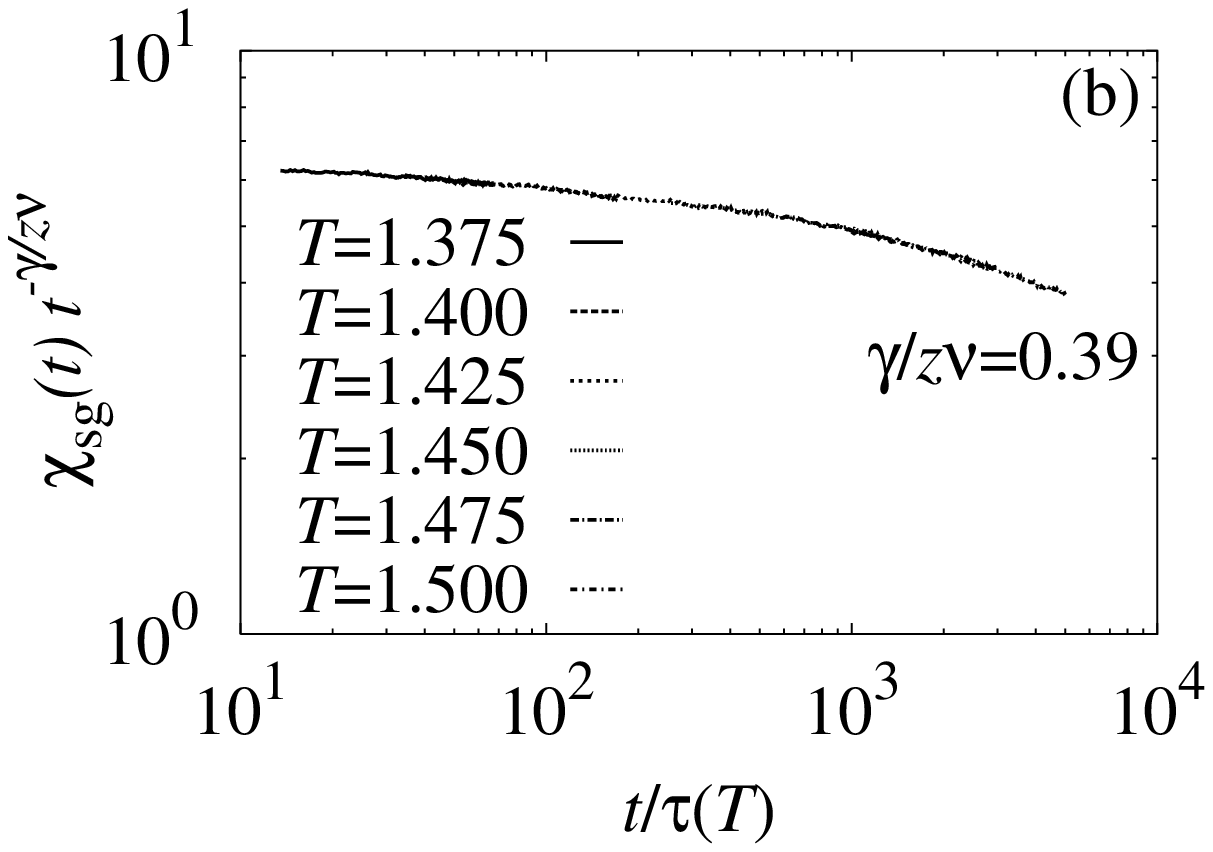}

\includegraphics[width=7.5cm]{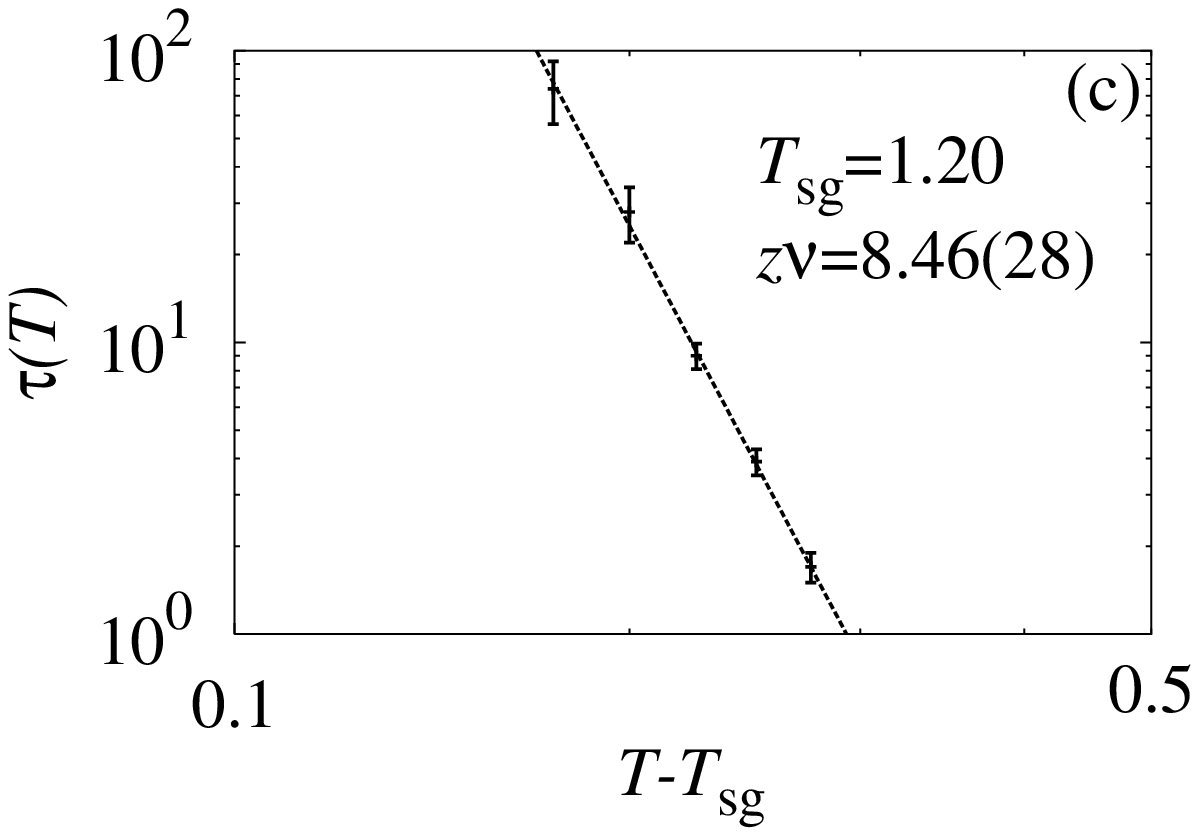}
\caption[]{(a) The NER of $\chi_\mathrm{sg}$ of the Ising model
at high temperatures.
(b)
The finite-time scaling plot for a choice of $\gamma/z\nu=0.39$.
We obtain $\tau(T)$  and $\gamma/z\nu$ so that
this scaling plot is good.
The scaling is also possible for $\gamma/z\nu=0.38\sim 0.40$.
(c)
The least-squares fitting of $\tau(T)$
supposing $\tau(T)\propto |T-T_\mathrm{sg}|^{-z\nu}$.}
\label{fig:isingsg}
\end{figure}

\Fref{fig:isingsg} shows an analysis to determine the critical
temperature and the exponent.
The simulation is performed just to check that our method gives 
results consistent with previous investigations
\cite{ogielski,bhatt,kawashimayoung,palassini,maricampbell}.
Therefore, the system size is very small ($L=19$) and the time range is very
short.
Finite-size effects are found to appear for $t> 5000$ by comparing with
results of $L=29$.
Only data before this time are used in the scaling analysis.
\Fref{fig:isingsg} (b) is an example of finite-time scaling.
A choice of $\gamma/z\nu$ is possible for 
$\gamma/z\nu=0.38\sim 0.40$.
A set of the correlation time at each temperature
is estimated for each choice of this exponent.
Then, the critical temperature is obtained
as summarised in \tref{tab:isingtsg}.
As the exponent increases, $T_\mathrm{sg}$ decreases.
We ignore a result of $\gamma/z\nu=0.400$ which deviates a lot from the others.
Our estimates are 
\begin{equation} 
T_\mathrm{sg}=1.17(4), \gamma/z\nu=0.3875(75), z\nu=9.3(12), \gamma=3.7(5).
\end  {equation} 
\begin{table}[ht]
\centering
\caption {
A list of the critical temperature and an exponent $z\nu$ obtained by
the finite-time scaling analysis in the Ising model.
A ratio of exponents $\gamma/z\nu$ denotes the possible value
in the finite-time scaling.
The least-squares fitting errors are denoted by $\chi^2$.}
\begin{tabular}{l|rrrr}
$\gamma/z\nu$ & $T_\mathrm{sg}$ & $z\nu$ &$\gamma$& $\chi^2$ \\
\hline
0.400 & 1.05 & 12.63(18) &5.1(1)& 1.40\\
0.395 & 1.13 & 10.07(37) &4.0(2)& 0.66\\
0.390 & 1.20 &  8.46(28) &3.3(1)& 0.54 \\
0.385 & 1.22 &  8.67( 9) &3.3(0)& 1.49\\
0.380 & 1.21 &  9.82( 9) &3.7(0)& 1.84
\end  {tabular}
\label{tab:isingtsg}
\end  {table}

\begin{figure}[ht]
\centering
\includegraphics[width=7.5cm]{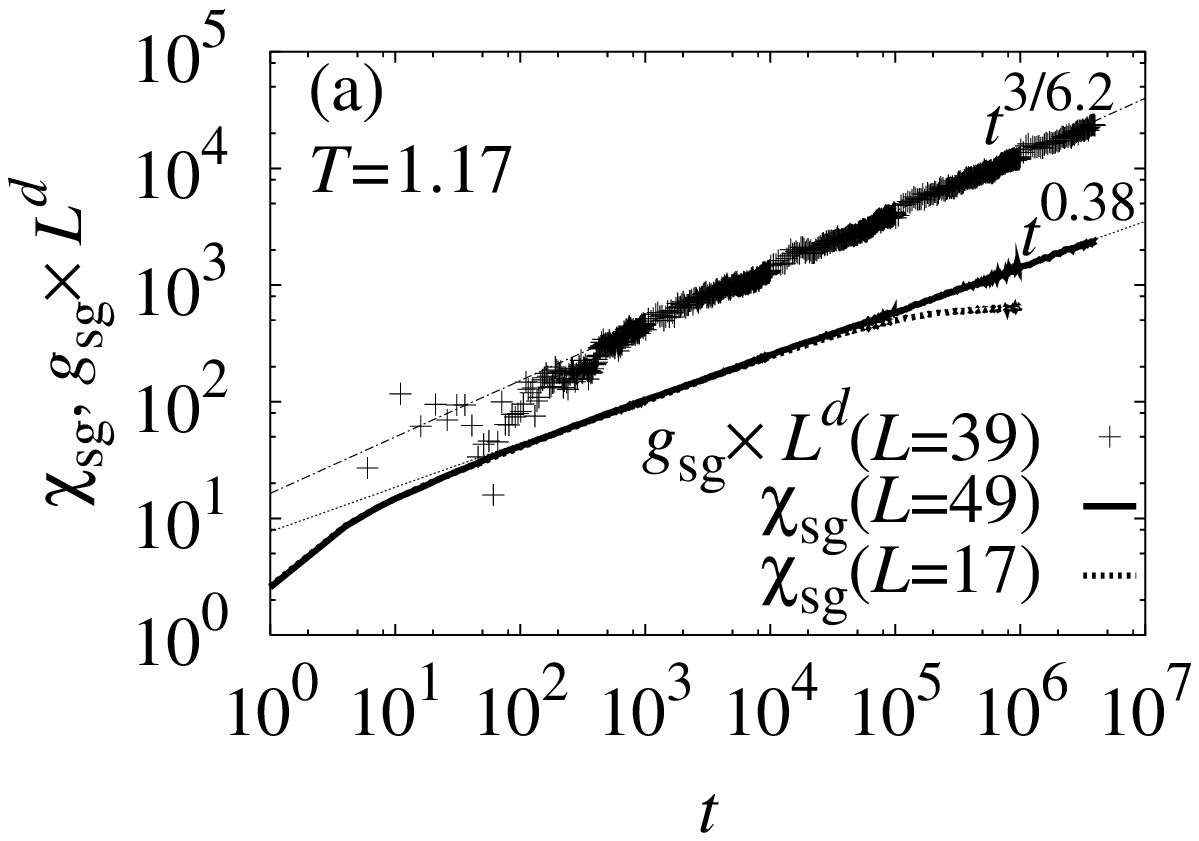}
\includegraphics[width=7.5cm]{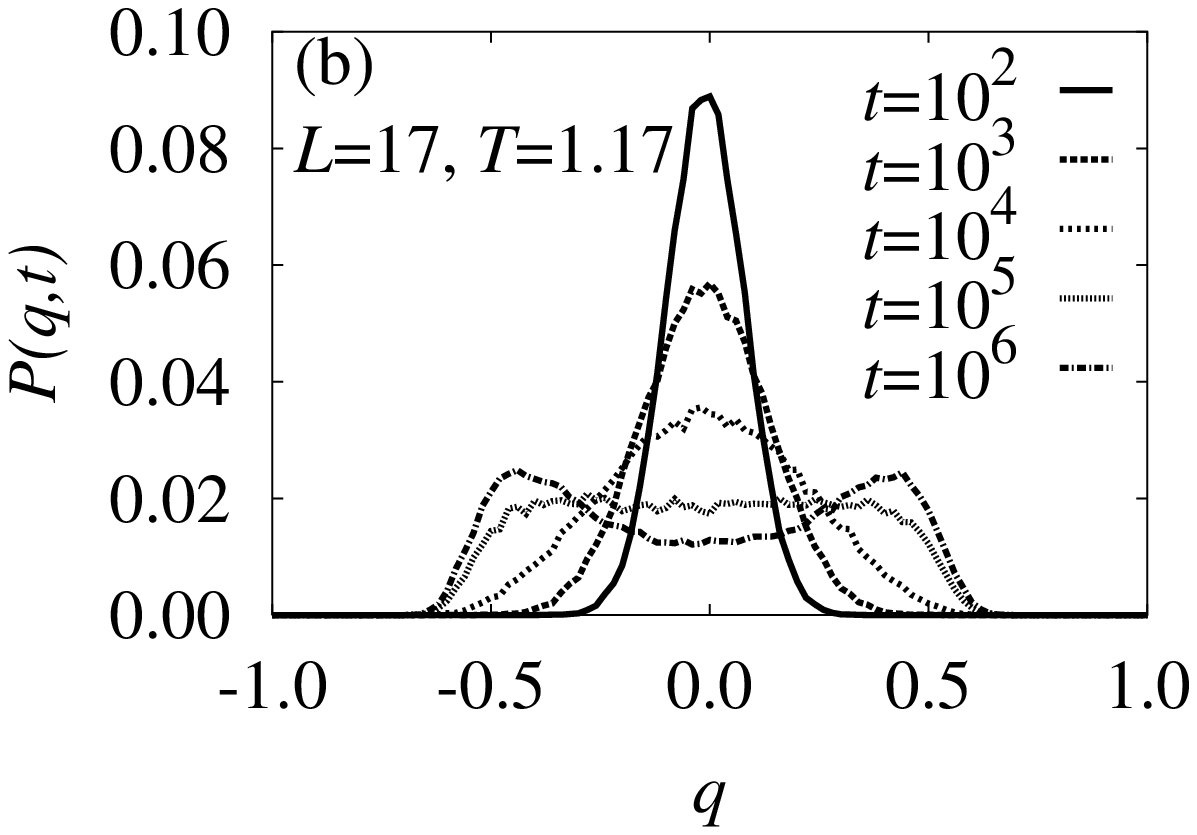}

\includegraphics[width=7.5cm]{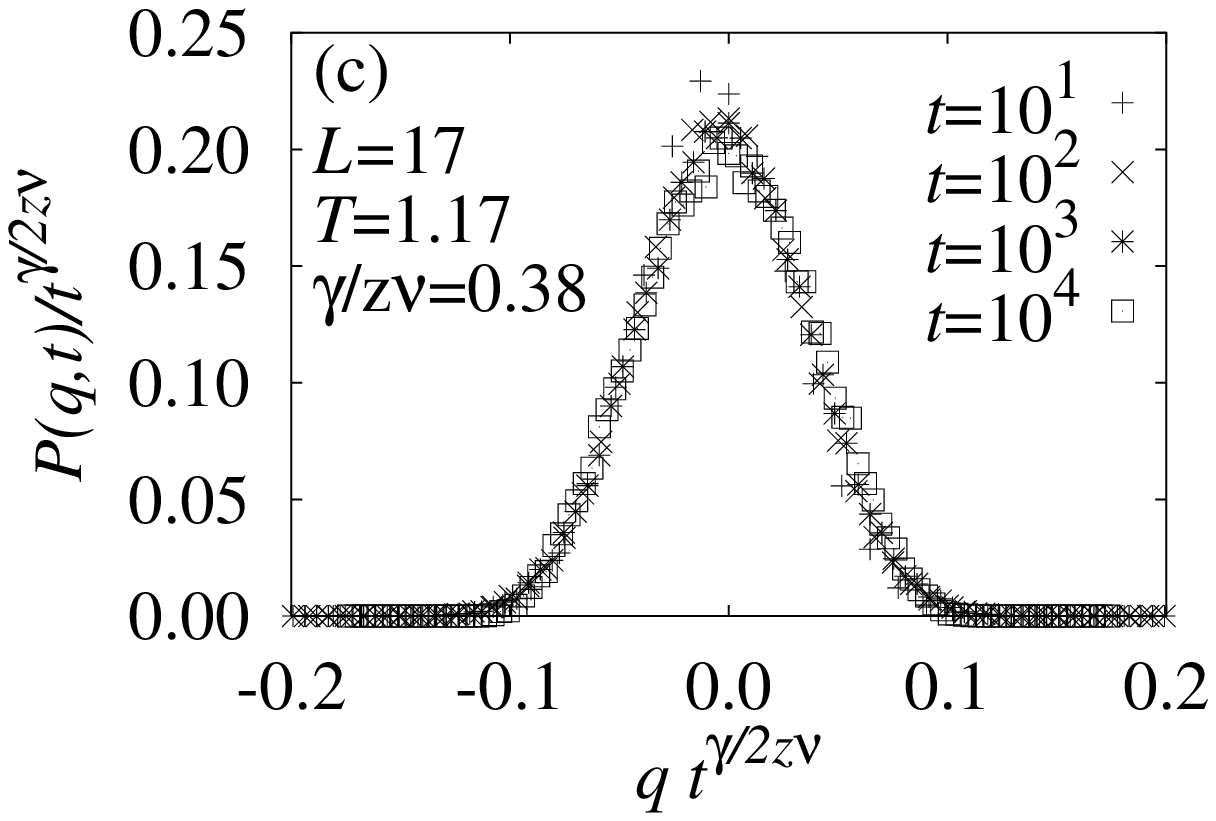}
\includegraphics[width=7.5cm]{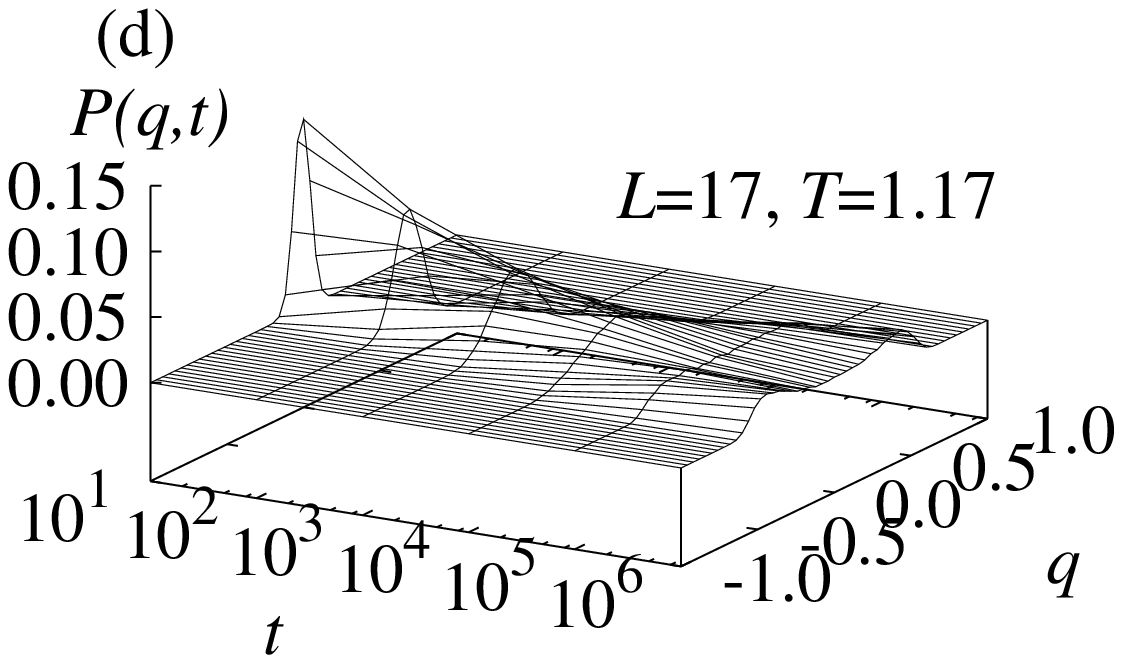}
\caption[]{(a) The NER of $\chi_\mathrm{sg}$ and $g_\mathrm{sg}\times L^d$
of the Ising model at $T_\mathrm{sg}=1.17$. Two lines, $t^{\gamma/z\nu}$ with
$\gamma/z\nu=0.38$ and $t^{d/z}$ with $z=6.2$, are guides for eyes.
(b)
The NER of the distribution function $P(q,t)$ at $T_\mathrm{sg}$ for $L=17$.
The shape changes from the single-peaked to the double-peaked when the
size effect of $\chi_\mathrm{sg}$ appears at $t=10^5$.
(c)
The finite-time scaling plot of $P(q,t)$.
(d)
A three-dimensional plot of $P(q,t)$.}
\label{fig:pofq}
\end{figure}

The results of the finite-time scaling analysis are checked by the raw
NER data at the obtained critical temperature.
\Fref{fig:pofq} (a) shows relaxation data of $\chi_\mathrm{sg}$ and 
$g_\mathrm{sg} \times L^d$. 
The SG susceptibility diverges algebraically with an exponent
$\gamma/z\nu=0.38$  that is consistent with the scaling result
$\gamma/z\nu=0.3875(75)$.
The critical relaxation process begins around 
$t\sim 100$ and seems to continue to infinity.
The Binder parameter also shows diverging behaviour with $t^{d/z}$,
from which we obtain the dynamic exponent $z=6.2(2)$.
Then, an exponent $\nu$ is estimated as $\nu=1.5(3)$.
A ratio of the critical exponents $\gamma/\nu=2.4(1)$.
The obtained results are consistent with previous numerical investigations 
\cite{ogielski,bhatt,kawashimayoung,palassini,maricampbell} 
and the corresponding experimental results
\cite{isingsgexp} as summarised in \tref{tab:explist}.
Since the lattice size and the time range are insufficient, the final
numerical results have large error bars.
As discussed in the previous section the critical temperature may be
underestimated by using a small lattice.
We plan to estimate them with high accuracy by large-scale NER analyses.

A time evolution of the distribution function of the overlap $P(q, t)$ 
at $T=T_\mathrm{sg}=1.17$ is shown in \fref{fig:pofq}(b).
The system size $L=17$.
It exhibits a single Gaussian form with a peak at $q=0$
before the size effect of $\chi_\mathrm{sg}$ appears at $t=10^5$
as shown in \fref{fig:pofq}(a).
As the time increases, the width of the distribution grows 
in accordance with the divergence of the spin-glass susceptibility.
It is possible to scale $P(q, t)/t^{\gamma/2z\nu}$ plotted versus
$qt^{\gamma/2z\nu}$ for various time steps from $t=10$ to $t=10^4$
(\fref{fig:pofq} (c)).
The critical exponent $\gamma/z\nu$ is also consistent with the finite-time
scaling of $\chi_\mathrm{sg}$.
The scaled data deviate a little for $t=10$ because the time is just before the
relaxation of $\chi_\mathrm{sg}$ 
reaches the critical relaxation region as shown in 
\fref{fig:pofq} (a).
The distribution changes its shape to having two peaks at $\pm q_\mathrm{eq}$
after the finite-size effect appears.
The shape is flat at this crossover time.

It is found that
the NER function knows the critical phenomenon from its very early time steps:
$t=10\sim 100$.
The NER method is now clearly shown to be applicable 
to the spin glass phenomenon.

\subsection{Heisenberg model}
\label{sec:result.heisen}

\begin{figure}[ht]
\centering
\includegraphics[width=7.5cm]{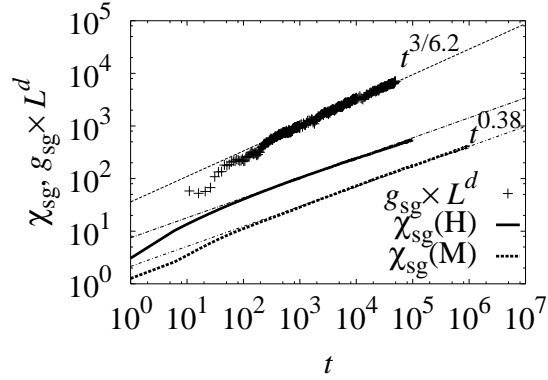}
\caption[]{The NER of $\chi_\mathrm{sg}$ and $g_\mathrm{sg}\times L^d$
of the Heisenberg model at $T_\mathrm{sg}=0.21$. 
Lattice sizes are $L=39$ for $g_\mathrm{sg}\times L^d$,
$L=89$ for $\chi_\mathrm{sg}$(H) and
$L=59$ for $\chi_\mathrm{sg}$(M).
Indices (H) and (M) denote the heat-bath updated and the Metropolis updated,
respectively.  Lines, $t^{\gamma/z\nu}$ with $\gamma/z\nu=0.38$ and
$t^{d/z}$ with $z=6.2$, are guides for eyes.}
\label{fig:heisen}
\end{figure}

We apply the same analysis performed in the Ising model to the Heisenberg model.
Finite-time scaling results have already been shown briefly in 
Ref. \cite{totasg1} and the detailed analysis will be reported elsewhere.
The system size is $L=59$ and the time scale is 70000 Monte Carlo steps. 
Typical numbers of bond configurations are same as given in 
\tref{tab:samplist}.
The finite-time scaling results are
\begin{equation}
T_\mathrm{sg}=0.20(2), \gamma/z\nu=0.39(5), z\nu=4.8(10), \gamma=1.9(5).
\end  {equation}
These results are checked by the raw NER at $T=0.21$ 
as shown in \fref{fig:heisen}.
The SG susceptibility diverges algebraically with an exponent
$\gamma/z\nu=0.38$, which is consistent with the result of the finite-time
scaling.
We performed simulations of both Metropolis update and heat-bath update.
The Metropolis result denoted by (M) and the heat-bath result denoted by (H)
exhibit the same critical behaviour, while the amplitudes are different by
a factor of 3.5.
NER behaviours are independent from the update algorithm.
The consistency supports the criticality of the SG order at
this temperature.
It is noted that the critical divergence begins at a very early time:
$t\sim 100$.

The Binder parameter exhibits a critical divergence $t^{d/z}$  with $z=6.2(5)$.
The value is consistent with that in the Ising model.
Since $z\nu$ is obtained by finite-time scaling, $\nu$ is estimated as
$\nu=0.8(2)$.
A ratio of the critical exponent $\gamma/\nu= 2.3(3)$.
The results are compared with an experimental result \cite{heisenexp} in 
\tref{tab:explist}.
They are not inconsistent.

\subsection{XY model}

It has been considered that there is no SG transition in this model
\cite{xynosg1,xynosg2}.
Only the CG transition with respect to the vector chirality
is considered to occur \cite{xychiral1,xychiral2,xychiral3}.
However, a possibility of the SG transition has recently been identified
by several investigations \cite{xyoccurs1,xyoccurs2,xyoccurs3}.
We applied the NER analysis on this model and our result supports
the latter conclusion: the SG transition occurs.

Our finite-time scaling analysis on the XY model is not yet conclusive
in regard to whether the SG transition
and the CG transition occur at the same temperature or not.
A system size ($L=39$), Monte Carlo steps ($10^5$) and a temperature range
used in the scaling analysis are insufficient to extract a conclusion.
However, as we increase the size and the steps, both critical temperatures
seem to approach each other: $T_\mathrm{sg}$ increase from low and
$T_\mathrm{cg}$ decrease from high.
Therefore, we consider that both transitions occur simultaneously.
Investigations are now being carried out and the details will be reported
elsewhere.
What has now been made clear is that both transitions occur in a 
temperature range of $ 0.4 < T < 0.46$.
In this paper we do not examine the issue of simultaneous transition
but focus on the existence of the SG transition.

\begin{figure}[ht]
\centering
\includegraphics[width=7.5cm]{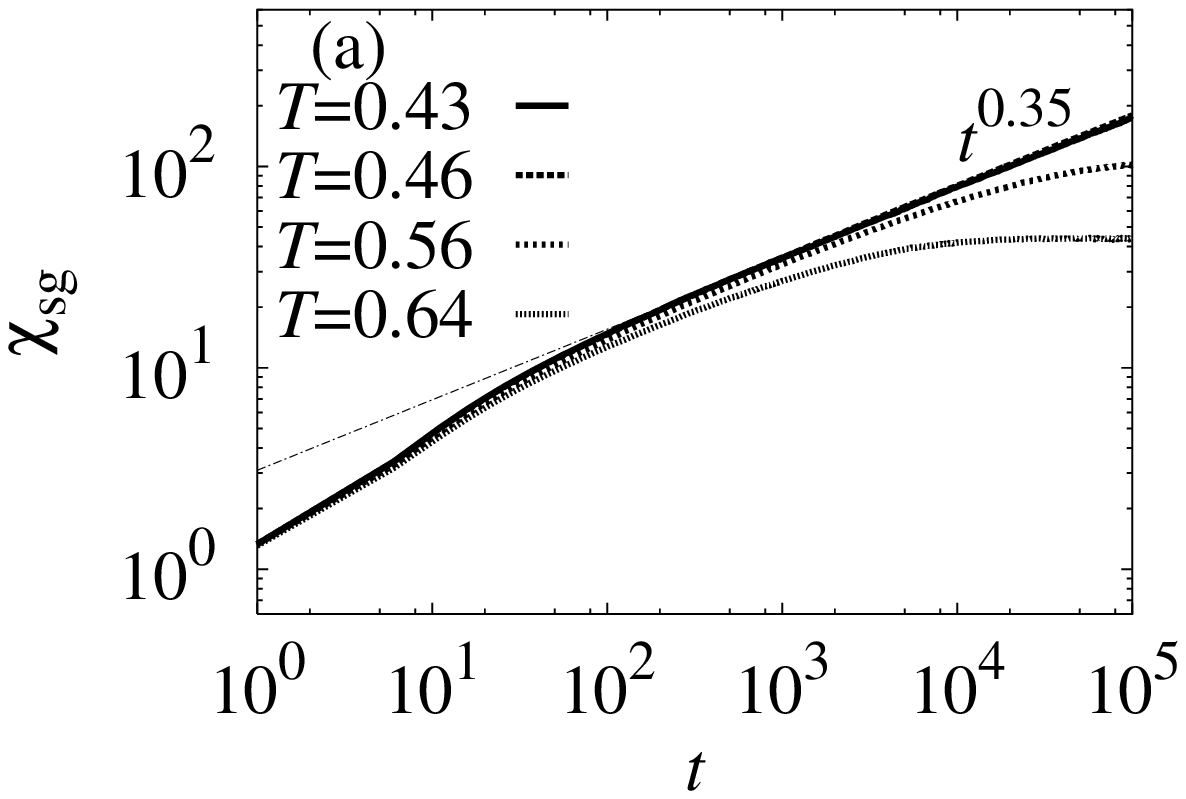}
\includegraphics[width=7.5cm]{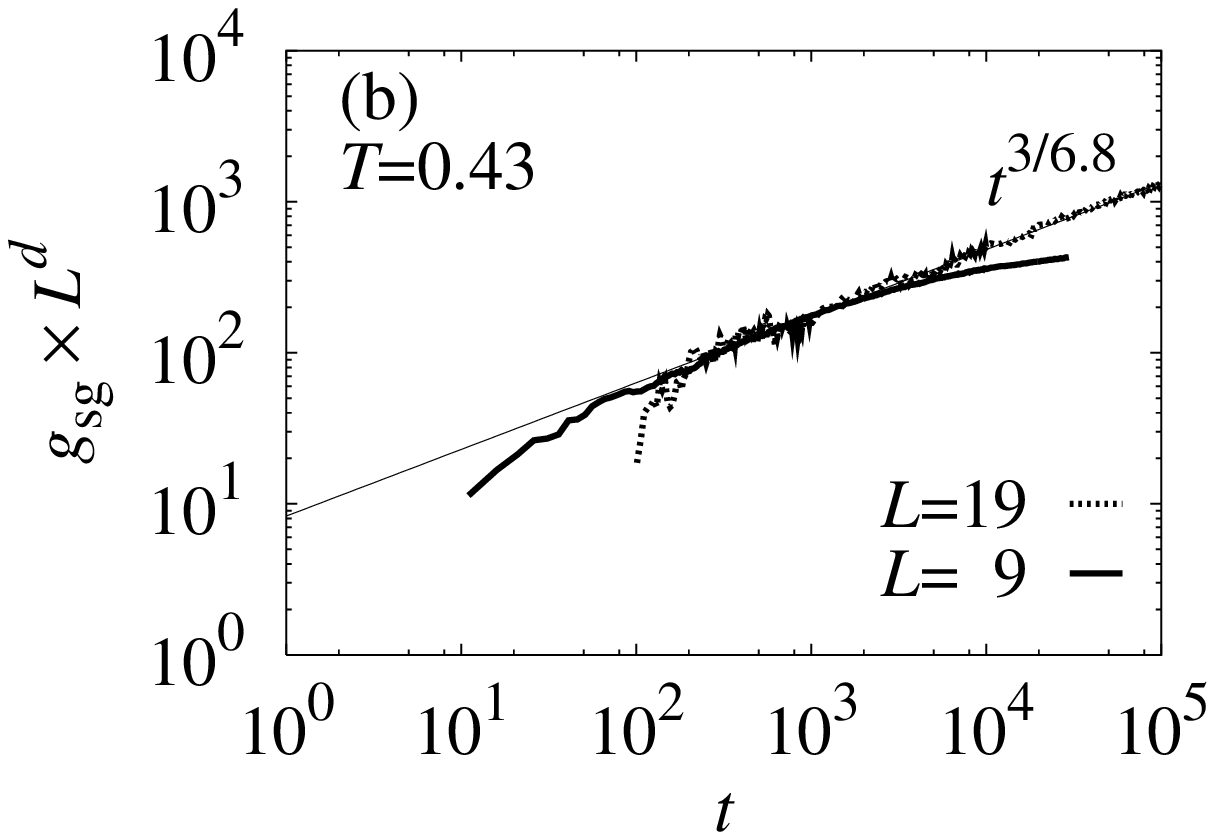}
\caption[]{(a) The NER of $\chi_\mathrm{sg}$ in the XY model above the critical
temperature $T_\mathrm{sg}\sim 0.43$. 
Lattice size is $L=39$.
A line $t^{\gamma/z\nu}$ with $\gamma/z\nu=0.35$ is a guide for eyes.
(b) The NER of the Binder parameter multiplied by $L^d$.
A line $t^{d/z}$ with $z=6.8$ is a guide for eyes.}
\label{fig:xy}
\end{figure}

\Fref{fig:xy}(a) shows raw NER plots of $\chi_\mathrm{sg}$ and 
$\chi_\mathrm{cg}$ near and above the critical temperature.
There is no difference in $\chi_\mathrm{sg}$ between $T=0.43$ and $T=0.46$
within the present time steps.
They exhibit a critical divergence with the same exponent and amplitude.
The SG transition is considered to occur near $T=0.43$.
From the slope we obtain an exponent $\gamma/z\nu=0.35$.
Note that this value is a little smaller than that of the Ising model and
the Heisenberg model ($\gamma/z\nu \sim 0.38$).
The NER of the Binder parameter is shown in \fref{fig:xy}(b).
It exhibits a critical divergence with an exponent $d/z$ with
$z=6.8(5)$, which is also a little larger than that of the other models.
However, we obtain a ratio of the critical exponents
$\gamma/\nu=2.4(2)$, which is consistent with the other models.

\subsection{Weak universality}

The SG transition occurs in all models as shown in the preceding subsections.
A ratio of the critical exponents $\gamma/\nu$ takes a common value 
around 2.4.
Therefore, there is a possibility of weak universality among these
transitions.
Not only a value of $\gamma/\nu$ but the NER functions themselves
suggest that the transitions are qualitatively equivalent. 

\begin{figure}[ht]
\centering
\includegraphics[width=7.5cm]{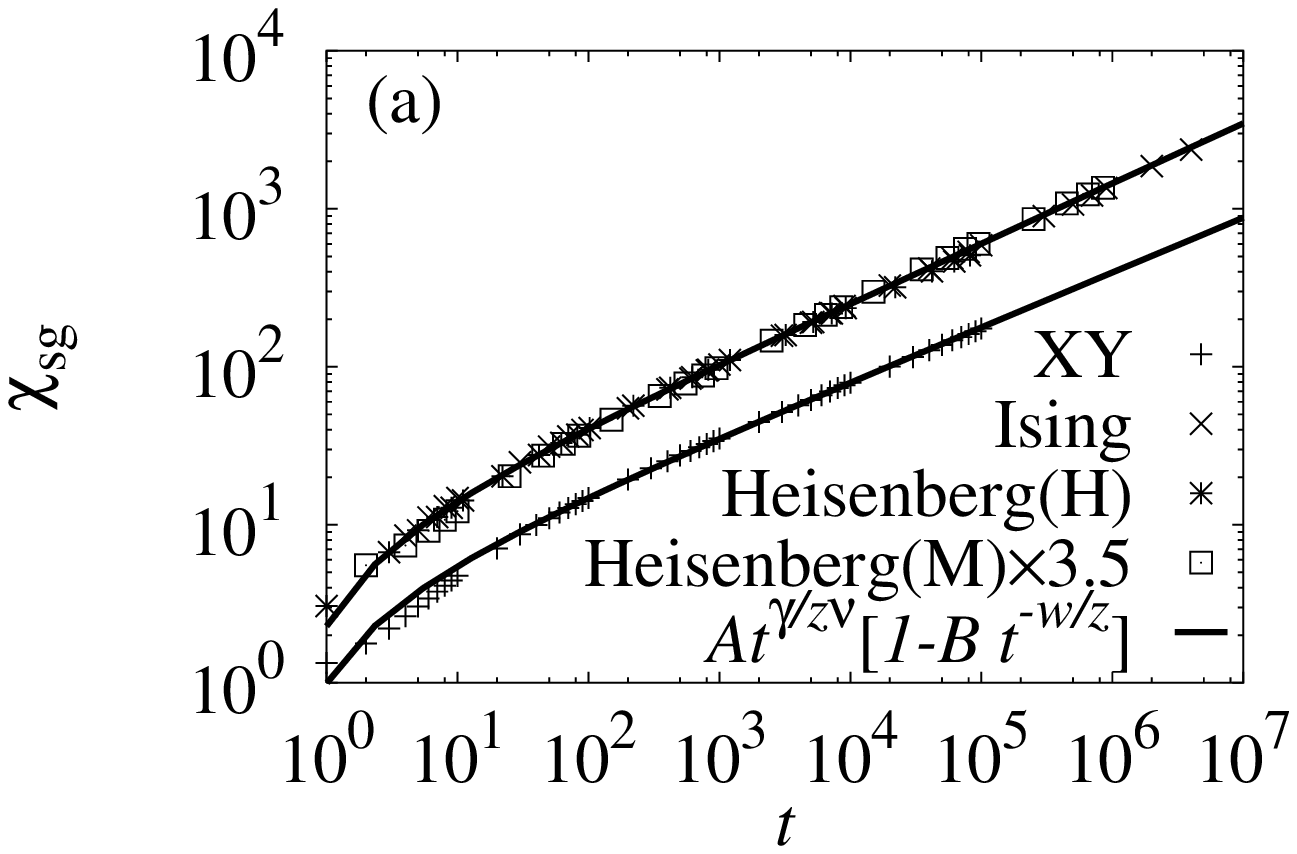}
\includegraphics[width=7.5cm]{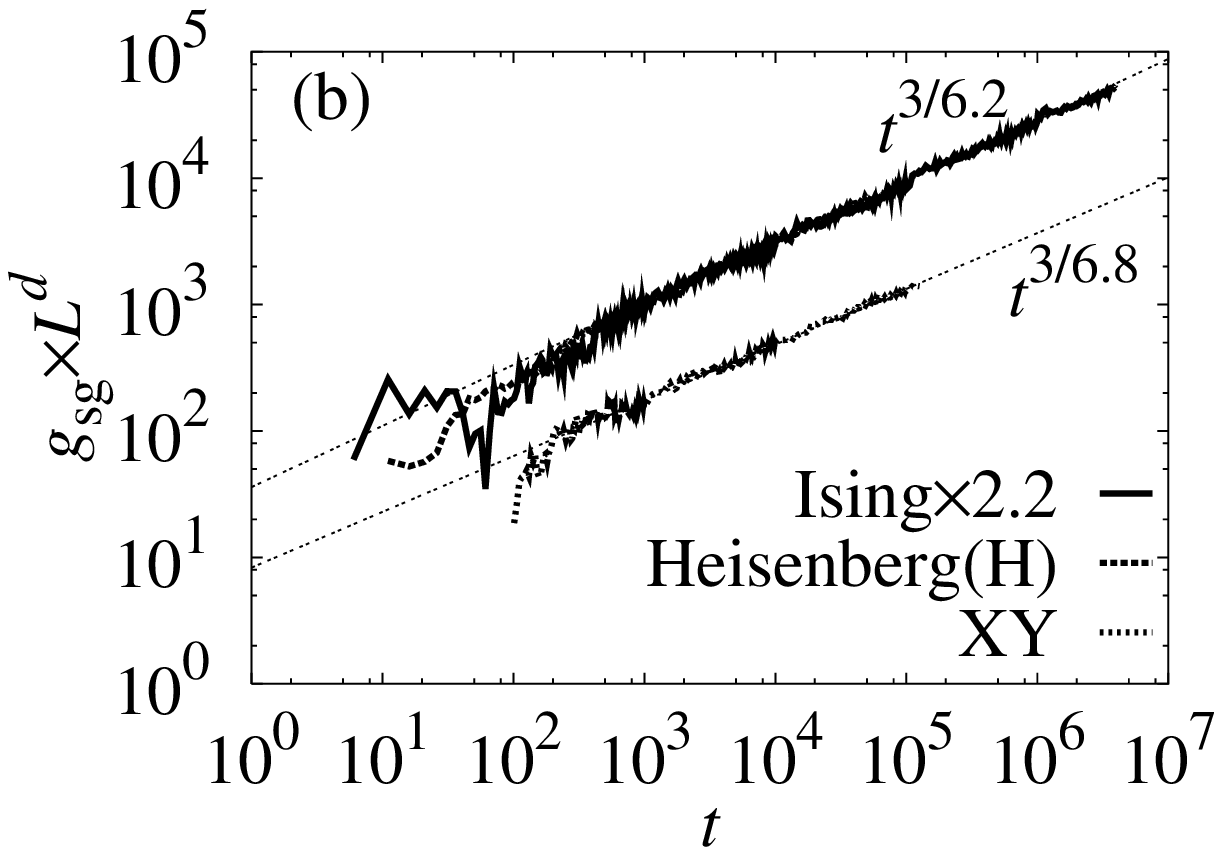}
\caption[]{(a) NER plots of the $\chi_\mathrm{sg}$ at $T_\mathrm{sg}$.
Correction-to-scaling fittings are depicted by bold lines with
$\gamma/\nu=2.356$, $w=3$, $z=6.2$ for the Ising/Heisenberg
model and $z=6.8$ for the XY model.
NER functions except the XY model are indistinguishable.
The data of the Heisenberg model with the Metropolis update are multiplied
by 3.5.
(b) NER plots of the Binder parameter multiplied by $L^d$.
The relaxation function of the Ising model multiplied by 2.2
coincides with that of the Heisenberg model with the heat-bath update.}
\label{fig:weakuniv}
\end{figure}

\Fref{fig:weakuniv} (a) shows the NER functions of $\chi_\mathrm{sg}$
at the critical temperature for all the models treated in this paper.
The data of the Heisenberg model with the Metropolis update are multiplied by
a factor 3.5 in order to compare with a result of the Ising model and
that of the Heisenberg model with the heat-bath update.
These three NER functions are not distinguishable.
If we take into account a correction-to-scaling term, 
the relaxation functions can be fitted from the first few steps
(Bold lines in \fref{fig:weakuniv} (a))
by an expression \cite{maricampbell}:
\begin{equation}
At^{\gamma/z\nu}[1-B t^{-w/z}].
\end{equation}
Here, exponents of the leading term are set $\gamma/\nu=2.356$ and $z=6.2$.
The correction-to-scaling exponent $w=3$.
Coefficient constants are $A=7.6$ and $B=0.7$.
The same expression also fits the NER function of the XY model,
but with the dynamic exponent $z=6.8$ and a constant $A=3.3$.
The NER functions of the Binder parameter are shown in 
\fref{fig:weakuniv} (b).
If we multiply the result of the Ising model by a factor 2.2, 
it is indistinguishable from the curve of the Heisenberg model.

\section{Summary}
\label{sec:summary}

By applying the nonequilibrium relaxation method
it has been made clear that the $\pm J$ models
in three dimensions undergo finite-temperature spin-glass transitions.
There is a possibility that these models belong to the same weak universality
class with a ratio of the critical exponents $\gamma/\nu \sim 2.4$.
We compare our results with other numerical results and the experimental
results in \tref{tab:explist}.
They agree well within the numerical errors.
Since the error bars are rather large at present,
further efforts to improve precision are necessary in order to prove
weak universality.

The spin-glass transition of the Heisenberg model is found to be very similar
to that of the Ising model.
The relaxation functions of $\chi_\mathrm{sg}$ and $g_\mathrm{sg}$ and values 
of a ratio of the exponents $\gamma/\nu$ and the dynamic exponent $z$ are
consistent between the two models.
If one considers that the spin-glass transition occurs in the Ising model, 
it may be thought that it occurs in the Heisenberg model in the same accuracy.
Only the dynamic exponent of the XY model differs from the other models.
Spin-glass transition and weak universality in models with
Gaussian bond distributions is a problem to be checked in future work.
The type of bond distributions may be important.

The NER method has been shown to be particularly effective 
in the spin-glass study.
Critical behaviour is observed from very early time steps
even though it takes a very long time to achieve the equilibrium states.
What is long is the nonequilibrium relaxation process after a short initial
relaxation before the final equilibrium relaxation.
This long process is discarded in conventional simulations, while it is
utilised in the NER method.
This is one reason why the NER method is advantageous in this system.
Applications to various complex systems with slow dynamics
are fruitful \cite{totaner1,totaner2}.

\begin{table}[ht]
\centering
\caption{Estimates of $T_\mathrm{sg}$, $\gamma$, 
$\nu$, $\gamma/\nu$ and $z$ in the $\pm J$ models in three dimensions.}
\begin{tabular}{llllll}
\hline
& $T_\mathrm{sg}$ & $\gamma$~ &  $\nu$ & $\gamma/\nu$ &$z$\\
\hline
Ising SG\\
Present work
& $1.17(4)$ & $3.6(6)$ & $1.5(3)$ &  $2.4(1)$& $6.2(2)$\\
Ref. \cite{kawashimayoung}
& $1.11(4)$ & 4.0(8) & $1.7(4)$ &  2.35(5) & \\
Ref. \cite{maricampbell}
& $1.195(15)$ & $2.95(30)$ & $1.35(10)$ & $2.225(25)$ & 5.65(15) \\
Experiment\cite{isingsgexp}
& & $4.0(3)$ & $\sim 1.7 $ &$\sim 2.4$ & \\
\hline
Heisenberg SG\\
Present work 
& $0.20(2)$ & $1.9(5)$ & $0.8(2)$ & $2.3(3)$&$6.2(5)$\\
Ref. \cite{matsubara3}
& 0.18(1) & 2.0(2) & 0.97(5) & 2.1(1)& \\
Experiment\cite{heisenexp}
& & $2.3(4)$ & 1.25(25) &2.0(7) & \\
\hline
XY SG\\
Present work & 0.43(3) &  &  & $2.4(2)$&$6.8(5)$\\
\hline
\end{tabular}
\label{tab:explist}
\end{table}

\ack%nowledgement
The authors would like to thank Professor Fumitaka Matsubara 
for guiding them in the spin-glass study and for their fruitful discussions.
The author TN also thanks
Professor Nobuyasu Ito and Professor Yasumasa Kanada for providing him with
a fast random number generator RNDTIK.
Computations were partly done at the Supercomputer Center, 
ISSP, The University of Tokyo.

\end{document}